\documentclass[aps,twocolumn,nofootinbib]{revtex4}
\usepackage[utf8]{inputenc}
 \usepackage{amsmath,amssymb}
 \usepackage{hyperref}
\usepackage{color}
\begin{document}

\title{An Effective Framework for Entropy Modifications via Stochastic Metric Fluctuations}
\author{Amir A. Khodahami \footnote{a.khodahami@shirazu.ac.ir} and Ahmad Sheykhi \footnote{asheykhi@shirazu.ac.ir}}
\affiliation{Department of Physics, College of Science, Shiraz
University, Shiraz 71454, Iran \\ Biruni Observatory, College of
Science, Shiraz University, Shiraz 71454, Iran}
\begin{abstract}
Deviations from the area law of horizon entropy are often modeled phenomenologically through modified Friedmann equations in cosmology. In this work, we explore an effective stochastic interpretation of such modifications by considering a Friedmann--Robertson--Walker (FRW) universe perturbed by a conformal, time-dependent noise factor with vanishing mean. Averaging the Einstein equations to second order in the fluctuation amplitude yields an effective correction to the Friedmann equation. We show that, in the spatially flat case with vanishing cosmological constant, this correction can be effectively matched to the general form induced by a deformed entropy-area relation. Within this phenomenological framework, the noise variance can be suitably chosen to match the functional forms associated with several generalized entropy models, including R\'enyi, (dual) Kaniadakis, Barrow, logarithmic, and MOND-inspired hypergeometric entropies. The present analysis therefore provides an effective correspondence between stochastic metric fluctuations and entropy deformations.
\end{abstract}
 \maketitle

 \section{Introduction}
Black hole thermodynamics suggests that gravitational field
equations originate from underlying thermodynamic principles.
    This idea is rooted in the deep correspondence between
    thermodynamic quantities, such as entropy and temperature,
    and geometric properties of horizons, including
    area and surface gravity. A central result in this direction
    is Jacobson's derivation of Einstein's equations from the
    fundamental relation $\delta Q = T \delta S$ applied to
    local horizons \cite{jacobson1995}. This thermodynamic derivation has since been extended to modified theories of gravity,
    including $f(R)$ gravity, Gauss-Bonnet gravity,
    scalar-tensor gravity, and more general Lovelock models
    \cite{eling2006,padmanabhan2005,paranjape2006,akbar2006,padmanabhan2010a}.

    In cosmology, the thermodynamics--gravity correspondence
    becomes particularly transparent. For an FRW Universe, the first Friedmann
    equation evaluated at the apparent horizon can be obtained
    from the first law of thermodynamics,
    $dE = T dS + W dV$, and conversely \cite{wang2006,cai2005,akbar2007,sheykhi2007a,sheykhi2010a,sheykhi2010b}.
    This interplay between boundary thermodynamics and
    bulk gravitational dynamics provides a concrete realization
    of holographic principles in cosmological settings.

Additional support for the emergent perspective of gravity comes
from statistical-mechanical considerations. Verlinde's entropic
gravity proposal \cite{verlinde2011} showed that
    Newtonian gravity can be interpreted as an entropic force
    arising from holographic and equipartition principles.
    More generally, gravity may be viewed as an emergent
    phenomenon associated with the microscopic degrees of
    freedom of spacetime \cite{cai2010,sheykhi2010,visser2011,sheykhi2013}. In this spirit,
    Padmanabhan proposed that cosmic expansion itself can
    be understood as the emergence of space driven by the
    difference between boundary and bulk degrees of freedom,
    from which the Friedmann equations naturally follow
    \cite{padmanabhan2012,cai2012,yang2012,sheykhi2013a}.

Recent developments in the thermodynamics--gravity correspondence
suggest that deviations from the standard Bekenstein-Hawking
entropy area relation of cosmological horizons can lead to
modified Friedmann equations governing cosmic expansion
\cite{sheykhi2010,nojiri2019,nojiri2022,odintsov2023,cai2008,sheykhi2018,saridakis2020,sheykhi2021,sheykhi2023}.
Such deviations are typically motivated by quantum-gravitational
effects or nonextensive statistical frameworks, in which the
horizon entropy acquires generalized forms. Prominent examples
include entropy constructions based on Tsallis \cite{tsallis1988},
Barrow \cite{barrow2020}, and Kaniadakis statistics
\cite{kaniadakis2002,kaniadakis2005}. These approaches introduce
deformation parameters that quantify departures from the
conventional entropy-area law.

Within the Tsallis framework, the nonextensive nature of entropy
leads to power-law modifications of the Friedmann equations
\cite{sheykhi2018}. Barrow entropy, inspired by possible fractal
structures of the horizon arising from quantum-gravitational
spacetime foam effects, incorporates a deformation parameter that
further alters the expansion dynamics
\cite{saridakis2020,sheykhi2021,sheykhi2023}. Similarly,
Kaniadakis statistics, which generalizes relativistic phase-space
distributions, results in modified cosmological evolution
equations
\cite{lymperis2021,hernandez-almada2022,drepanou2022,zangeneh2024,sheykhi2024}.
Since each generalized entropy formalism yields distinct
corrections to the Friedmann equations, they can significantly
affect early-Universe processes, including the neutron-to-proton
freeze-out ratio, the deuterium bottleneck, and the resulting
primordial abundances of light elements.

In this work, we investigate whether stochastic fluctuations of the spacetime metric can reproduce, at the level of effective Friedmann dynamics, correction terms analogous to those appearing in generalized entropy models. Specifically, we consider conformal noise with a vanishing coarse-grained average on an FRW spacetime, ensuring that the mean geometry remains FRW. In the perturbative regime considered here, linear noise contributions vanish under averaging, while nonlinear terms in Einstein’s equations survive at second-order in the fluctuation amplitude, providing the leading non-vanishing contribution in this framework. The resulting Friedmann equation then acquires an effective correction term that can be matched, through an appropriate choice of the noise variance profile, to the forms associated with generalized entropy formalisms, without imposing a modified entropy law \emph{a priori}. Furthermore, this framework preserves the standard Einstein-Hilbert action, which distinguishes our derivation from studies that obtain modified entropy relations via fundamental action-level corrections \cite{Dehyadegari:2026PLB}. From this perspective, deviations from the standard area law may be viewed as an effective coarse-grained parametrization of unresolved stochastic spacetime degrees of freedom, with horizon thermodynamics providing a convenient language for capturing their macroscopic imprint. We do not attempt to provide a microscopic theory of metric noise; rather, our goal is to test whether a simple stochastic ansatz can account for the effective form of Friedmann-equation corrections associated with generalized horizon entropies. For the explicit stochastic derivation developed below, however, we restrict attention to the spatially flat case with vanishing cosmological constant, so the resulting correspondence should be understood only within this limited background.

It is important to clarify that the present approach differs
conceptually from the Einstein-Langevin framework of stochastic
semiclassical gravity. In the Einstein-Langevin equation,
stochasticity appears explicitly as a random source term in the
field equations, typically arising from quantum stress--tensor
fluctuations, so that one solves a genuinely stochastic
differential equation for the metric \cite{martin1999}. In
contrast, we do not modify Einstein's equations by adding an
external noise term. Instead, we consider a metric with stochastic fluctuations that satisfies the classical Einstein equations,
and then perform a coarse-graining over these fluctuations. 
The stochasticity therefore resides in the space of solutions 
rather than in the dynamical equations themselves.

This paper is organized as follows. In Sec.~\ref{Sec:Mod-Frid}, we
study the apparent horizon of an FRW Universe in the presence of
stochastic conformal fluctuations and obtain the corresponding effective modification of the Friedmann equation. In
Sec.~\ref{Sec:Spec-Cases}, we compare the induced correction term
with those obtained from well-known generalized entropy frameworks
and illustrate how specific noise statistics can match these effective entropy deformations. Finally, we summarize our findings and discuss the limitations and implications of this correspondence in Sec.~\ref{Sec:Concs}.
\section{Modified Friedmann Equations from Conformal Metric Noise}   \label{Sec:Mod-Frid}
The dynamics of an FRW Universe admit a thermodynamic
interpretation when evaluated at the apparent horizon. The FRW
metric may be written as
\begin{equation}
ds_{\mathrm{FRW}}^2 = h_{\mu\nu} dx^\mu dx^\nu + R^2
\left(d\theta^2 + \sin^2\theta d\phi^2\right),
\label{eq:frw-metric}
    \end{equation}
    where $R = a(t) r$ is the physical radius, $x^0=t$, $x^1=r$, and
    \begin{equation}
        h_{\mu\nu} = \mathrm{diag}\!\left(-1,\frac{a^2(t)}{1-k r^2}\right),
    \end{equation}
denotes the metric of the two-dimensional $(t,r)$ subspace. The
parameter $k = -1, 0, 1$ characterizes the spatial curvature of
the Universe, corresponding to open, flat, and closed geometries,
respectively. From a thermodynamic standpoint, the relevant causal
boundary in an FRW spacetime is the apparent horizon. Its radius
is given by
    \begin{equation}
        R_h = \left(H^2 + \frac{k}{a^2}\right)^{-1/2},
        \label{eq:ah_radius}
    \end{equation}
where $H = \dot a / a$ is the Hubble parameter and the overdot
denotes differentiation with respect to cosmic time. In the spatially flat case ($k=0$), this reduces to $R_h = 1/H$ for an expanding universe with $H>0$. For a homogeneous and isotropic Universe filled with a perfect fluid of energy density $\rho$ and pressure $p$, energy--momentum
conservation leads to the continuity equation
    \begin{equation}
        \dot{\rho} + 3H(\rho + p) = 0.
        \label{eq:continuity}
    \end{equation}
On the other hand, the $00$ component of Einstein's field
equations with cosmological constant $\Lambda$ yields the
Friedmann equation
    \begin{equation}
        \frac{8\pi G}{3}\rho + \frac{\Lambda}{3}
        =
        \frac{1}{R_h^2},
        \label{eq:friedmann}
    \end{equation}
where $R_h$ is the apparent horizon radius given by
Eq.~\eqref{eq:ah_radius}. It is well known that
Eq.~\eqref{eq:friedmann} can be recovered by combining the
continuity equation \eqref{eq:continuity} with the first law of
thermodynamics applied to the apparent horizon,
    \begin{equation}
        dE = T_h dS_h + W dV,
\end{equation}
provided that the horizon entropy obeys the Bekenstein-Hawking
area law $S_h = S_{BH} =A/(4G)$. Here $T_h$ denotes the horizon
temperature,
\begin{equation}
        T_h=\frac{\kappa_h}{2 \pi}
        =
        -\frac{1}{2 \pi R_h}\left(1-\frac{\dot{R_h}}{2 H R_h}\right).
    \end{equation}
Moreover, $V = 4\pi R_h^3/3$ denotes the volume enclosed by the
apparent horizon, and the work density $W$ is given by
    \begin{equation}
        W = -\frac{1}{2} T^{\mu\nu} h_{\mu\nu}
        = \frac{1}{2}(\rho - p).
    \end{equation}
This correspondence reveals that the standard Friedmann equation
encodes the thermodynamic properties of spacetime geometry
\cite{hayward1998,hayward1999,bak2000,akbar2007,cai2005}.

Motivated by this result, it is natural to investigate how
cosmological dynamics are modified when the entropy--area relation
is altered. This is done in
\cite{sheykhi2024,sheykhi2025,sheykhi2025a,sheykhi2025b} for
special cases. Following the derivations presented in these works,
one finds that a general deviation from the area law leads to a
modified Friedmann equation of the form
    \begin{equation}
            \frac{8\pi G}{3}\rho + \frac{\Lambda}{3}
        =
        \frac{1}{R_h^2}
        -\int \frac{2 dR_h}{R_h^3}  \frac{d\delta S_h}{dS_{BH}},
        \label{eq:Friedmann-general-mod}
    \end{equation}
where $\delta S_h$ denotes the deviation from the area law and
$S_{BH}=A/(4G)$.

Now we investigate how the thermodynamic description of an FRW
spacetime is affected by stochastic conformal fluctuations of the
metric. We assume that these fluctuations represent unresolved
gravitational degrees of freedom and have zero mean, so that the mean geometry coincides with the standard FRW
spacetime. Accordingly, we consider a perturbed metric of the form
    \begin{equation}
        ds^2_{\mathrm{CN}} = \left(1 + \epsilon b(t)\right) ds^2_{\mathrm{FRW}},
        \label{eq:CNmetric}
    \end{equation}
where $ds^2_{\mathrm{FRW}}$ is the FRW line element given in
Eq.~\eqref{eq:frw-metric}, $\epsilon$ is a small dimensionless
parameter controlling the amplitude of the fluctuations, and
$b(t)$ is a stochastic conformal noise. Restricting the perturbation to be purely conformal and to depend
only on time is a deliberate choice. This ensures that the
fundamental symmetries of the FRW Universe, in particular spatial
homogeneity and isotropy, are preserved, while still allowing for
nontrivial fluctuations in the gravitational sector. As a result,
the stochastic effects enter the dynamics only through nonlinear
contributions that survive upon coarse-graining, thereby
modifying the effective cosmological evolution without introducing
anisotropies or inhomogeneities.

The physical interpretation of the stochastic conformal factor $b(t)$
is tied to the separation of timescales between the underlying
microscopic fluctuations and the temporal resolution relevant for
cosmological observations. In the present framework, $b(t)$ is assumed
to describe rapidly varying unresolved gravitational fluctuations with
a characteristic correlation time $\tau$, whereas the cosmological
background evolves on the much longer Hubble timescale $H^{-1}$.
Moreover, any realistic cosmological inference is necessarily based on
observables averaged, explicitly or effectively, over a finite time
interval $\Delta t$ rather than on an instantaneous probing of the
microscopic metric realization. The relevant regime is therefore one in
which the fluctuation timescale is much shorter than the observational
or coarse-graining timescale,
\begin{equation}
	\tau \ll \Delta t,
	\qquad
	\tau \ll H^{-1},
	\label{eq:scale-separation}
\end{equation}
so that many microscopic oscillations occur within a single effective
measurement interval while the background expansion remains nearly
unchanged during that interval.

At the same time, one should not overinterpret the instantaneous
stochastic factor $b(t)$ as a directly measurable microscopic
observable. In general relativity, the detailed decomposition of a
metric fluctuation may depend on the choice of coordinates or time
parametrization, so a conformal perturbation written in this form is
not, by itself, the primary gauge-invariant object of physical
interest. The role of $b(t)$ in the present framework is therefore
effective rather than literal: it provides a convenient
phenomenological parametrization of unresolved fast geometric
fluctuations, while the physically relevant content resides in the
coarse-grained corrections that these fluctuations induce in the
averaged background dynamics.

Accordingly, the stochastic quantity $b(t)$ is not introduced as a
directly resolved observable in its instantaneous form. Rather, it
serves as a phenomenological parametrization of microscopic
fluctuations whose detailed time dependence is inaccessible at the
level of the effective cosmological description. What can influence the
macroscopic dynamics is only the cumulative imprint of these
fluctuations after averaging over times much larger than $\tau$. For a
generic quantity $X(t)$, this coarse-grained description may be
represented schematically by a finite-time average
\begin{equation}
	\langle X(t) \rangle
	=
	\frac{1}{\Delta t}
	\int_t^{t+\Delta t} X(t')\,dt',
	\qquad
	\Delta t \gg \tau,
	\label{eq:finite-time-average}
\end{equation}
which removes the rapidly oscillating part of $X$ while retaining its
systematic contribution to the slow cosmological evolution. In this
sense, the effective dynamics derived below should be understood as a
coarse-grained description appropriate to observers or probes that do
not resolve the microscopic fluctuation timescale.

We assume that the stochastic noise and its derivatives possess vanishing odd moments consistent with the zero-mean character of the noise
	\begin{equation}
		\label{eq:noise-ave-cond-1}
		\langle b(t) \rangle = \langle \dot b(t) \rangle = \langle \ddot b(t) \rangle = \langle b^3(t) \rangle=\cdots = 0.
	\end{equation}
	Consequently, the averaged line element reduces to the standard FRW form,
	\begin{equation}
		\langle ds^2_{\mathrm{CN}} \rangle = ds^2_{\mathrm{FRW}},
	\end{equation}
	by construction. The mixed correlation $\langle b(t)\dot b(t)\rangle$ is negligible in the coarse-grained description. Indeed,
\begin{equation}
	\label{eq:bdotb}
		\langle b\dot b \rangle
		= \frac{1}{\Delta t} \int_t^{t+\Delta t} b(t')\,\frac{db(t')}{dt'}\,dt'
		= \frac{1}{2\Delta t}\Bigl(b^2(t+\Delta t)-b^2(t)\Bigr).
\end{equation}
In our coarse-grained description, the averaging window is chosen such that $\Delta t \gg \tau$, where $\tau$ represents the noise correlation time. Consequently, the fluctuations at the boundaries $t$ and $t+\Delta t$ are effectively uncorrelated, meaning their contribution does not build up any systematic drift, so the boundary difference averages out. Even if one retains this boundary difference term, it is heavily suppressed compared to the dominant contributions of order $\tau^{-2}\langle b^2\rangle$ that appear later in the analysis, and can therefore be consistently neglected. By contrast, quadratic combinations such as $b^2$ and $\dot b^2$ are sign-stable in the averaged sense and therefore need not vanish. These second-order terms accumulate into a nontrivial correction to the Friedmann equations even though the mean of the fluctuation itself is zero.

Substituting the perturbed metric \eqref{eq:CNmetric} into Einstein's field equations, we henceforth restrict attention to the case of vanishing cosmological constant ($\Lambda=0$) and spatially flat geometry ($k=0$), in order to keep the stochastic calculation tractable. Accordingly, the explicit correspondence established below applies only within this restricted setting. Suppressing explicit time dependence for notational compactness, we obtain two independent equations. These correspond to the temporal ($00$) component and to the spatial components, which are identical for $11$, $22$, and $33$ by virtue of spatial homogeneity and isotropy. Expanding up to second order in $\epsilon$, the temporal component yields
    \begin{equation}
        \frac{8 \pi G}{3} \rho=\frac{\dot a^2}{a^2}+
        \frac{\epsilon\left(a \dot a \dot b-b \dot a^2\right)}{a^2}
        +\frac{ \epsilon^2\left(4 b^2  \dot a^2-8  a  b  \dot a  \dot b+a^2  \dot b^2\right)}{4  a^2},
    \end{equation}
    while the spatial components give
    \begin{align}
        8 \pi G  p&=-\frac{\dot a^2+2 a  \ddot a}{a^2}
        +\frac{\epsilon\left(b  \dot a^2-2 a  \dot a  \dot b+2 b  a  \ddot a  - a^2 \ddot b\right)}{a^2}
        \nonumber\\
        &+\frac{\epsilon^2\left(-4 b^2 \dot a^2+16 a b \dot a \dot b+3 a^2  \dot b^2-8 a  b^2 \ddot a+8 a^2 b  \ddot b\right)}{4 a^2}.
    \end{align}
    The conformal noise induces stochastic fluctuations in the energy density and pressure. In this framework, the Einstein equations are applied to the effective geometry with rapid local conformal fluctuations, and the corresponding energy density and pressure are those required for that geometry to satisfy the classical field equations. No additional independent stochastic stress tensor is introduced, and no external random source term is added to the right-hand side of Einstein's equations. Rather, the stochasticity is assumed to reside entirely in the metric sector, while the matter variables acquire their fluctuating character through their coupling to the noisy geometry. The physically relevant quantities in the effective cosmological description are therefore the coarse-grained averages $\langle \rho \rangle$ and $\langle p \rangle$. Upon performing the coarse-graining, all linear terms vanish under the zero-mean prescription of Eq.~\eqref{eq:noise-ave-cond-1}, whereas the mixed correlator $\langle b\dot b\rangle$ is neglected at leading order, in view of Eq.~\eqref{eq:bdotb} together with the discussion below it. The leading non-vanishing contributions therefore arise at second order in the fluctuation amplitude and its derivative. As a result, the coarse-grained Einstein equations take the form
    \begin{equation}
        \frac{8\pi G}{3} \langle \rho\rangle
        =
        H^2
        + \epsilon^2\left(H^2 \langle b^2\rangle
        +  \frac{\langle\dot b^2\rangle}{4}\right),
        \label{eq:modifiedFriedmann}
    \end{equation}
    and
    \begin{equation}
    8 \pi G  \langle p\rangle=-3 H^2-2 \dot H+\epsilon^2\left(\left(-3 H^2-2 \dot H\right)\langle b^2\rangle+\frac{3}{4}  \langle\dot b^2\rangle\right).
\end{equation}
These equations govern the coarse-grained cosmological dynamics,
with the quadratic terms encoding the leading imprint of the
underlying microscopic conformal fluctuations as systematic
corrections to the standard Friedmann equations. Given the rapidly oscillating noise characterized by Eq.~\eqref{eq:scale-separation}, one expects
\begin{equation}
    \langle \dot b^2 \rangle \sim \frac{\langle b^2 \rangle}{\tau^2}
    \gg H^2 \langle b^2 \rangle,
    \qquad
    \langle \dot b^2 \rangle \gg |\dot H|\langle b^2\rangle.
    \label{eq:noise-approx}
\end{equation}
Accordingly, the averaged equations may be approximated as
\begin{equation}
    \frac{8\pi G}{3} \langle \rho\rangle
    =
    H^2
    + \frac{\epsilon^2}{4}  \frac{\langle b^2 \rangle}{\tau^2},
    \label{eq:modifiedFriedmann-2}
\end{equation}
and
    \begin{equation}
    8 \pi G  \langle p\rangle=-3 H^2-2 \dot H - \frac{5\epsilon^2}{4}  \frac{\langle b^2 \rangle}{\tau^2}.
\end{equation}
Comparing Eq.~\eqref{eq:modifiedFriedmann} with
Eq.~\eqref{eq:Friedmann-general-mod} allows one to interpret the
stochastic correction as an effective modification of the horizon
entropy--area law in the spatially flat ($k=0$), $\Lambda=0$ case considered here. A convenient identification is
\begin{equation}
    \epsilon^2\left(H^2 \langle b^2\rangle
    +  \frac{\langle\dot b^2\rangle}{4}\right)\;\longleftrightarrow\;
    -\int \frac{2 dR_h}{R_h^3}  \frac{d\delta S_h}{dS_{BH}},
    \label{eq:match-noise}
\end{equation}
in the sense that both terms enter the Friedmann equation as
additive corrections to $H^2$. Under the approximations in Eq.~\eqref{eq:noise-approx}, this identification reduces to
\begin{equation}
    \frac{\epsilon^2}{4}\frac{\langle b^2\rangle}{\tau^2}
    \;\longleftrightarrow\;
    -\int \frac{2 dR_h}{R_h^3}\frac{d\delta S_h}{dS_{BH}}.
    \label{eq:match-noise-approx}
\end{equation}
The correspondence in Eqs.~\eqref{eq:match-noise} and \eqref{eq:match-noise-approx} is restricted to the sign-consistent sector, namely to entropy corrections whose effective contribution can be realized
by a real stochastic fluctuation with non-negative quadratic
amplitude. Entropic modifications requiring the opposite sign, some of which are highlighted in Sec.~\ref{Sec:Spec-Cases}, lie outside the scope of the present analysis.

Substituting the perturbed metric of Eq.~\eqref{eq:CNmetric} into the local conservation law $\nabla_\mu T^{\mu\nu}=0$ yields the modified continuity equation
\begin{equation}
	\dot \rho+3H(\rho+p)+\frac{3 \epsilon \left(\rho+p\right) \dot b}{2\left(1+ \epsilon b\right)}=0.
\end{equation}
This relation holds at the level of the fluctuating geometry, prior to coarse-graining, where the conformal noise is still explicitly present. The physically relevant cosmological description is obtained after averaging over timescales satisfying $\Delta t\gg\tau$, so that the rapidly oscillating contributions are washed out. Accordingly, up to $\mathcal{O}(\epsilon^2)$, the effective continuity equation reduces to
	\begin{equation}
		\langle\dot{\rho}\rangle
		+3H\bigl(\langle\rho\rangle+\langle p\rangle\bigr)=0.
	\end{equation}
	In this sense, the standard FRW conservation law is approximately preserved at the coarse-grained level. This property is important for the frameworks developed in \cite{sheykhi2025a,sheykhi2025b}, where the usual conservation equation plays a central role.
\section{Special Cases: Correspondences with Known Modified Entropies}
\label{Sec:Spec-Cases}
In the following, we illustrate how several widely studied generalized entropy models can be matched at the level of the effective Friedmann dynamics by appropriately specifying the statistical properties of the conformal noise, within the spatially flat ($k=0$), $\Lambda=0$ setting considered above. Throughout this section, we take the characteristic fluctuation timescale in Eq.~\eqref{eq:modifiedFriedmann-2} to be set by the Planck time, $\tau=t_{\rm Pl}$, corresponding to microscopic spacetime fluctuations at the quantum-gravity scale. This choice is motivated by the interpretation of horizon entropy as a measure
of the underlying fundamental degrees of freedom of spacetime, so
that entropy deformations naturally encode the coarse-grained
imprint of stochastic dynamics at the Planck scale. The resulting correspondences should be understood as phenomenological: they provide an effective mapping between entropy-corrected Friedmann equations and the stochastic conformal-noise framework, but do not constitute a microscopic derivation of the underlying noise kernel from a specific quantum-gravity theory. Accordingly, the constructions below are best viewed as effective testbeds for exploring how generalized entropy deformations may be encoded in horizon-dependent stochastic geometry.

This setup naturally accommodates a slow time dependence in the noise variance through its relation to the horizon scale. Within the coarse-graining procedure, rapid instantaneous fluctuations of the conformal factor are averaged out, whereas the slowly evolving cosmological background induces only a smooth temporal modulation of the variance. The resulting separation of scales makes these two features fully compatible: short-timescale stochastic fluctuations do not survive in the effective description, while a mild background-driven time dependence of the variance may consistently remain. This provides a physically coherent framework for relating spacetime fluctuations to scale-dependent entropy models.

A concrete realization of the above scale-separated picture is obtained by modeling \(b(t)\) as a zero-mean, smooth Gaussian colored fluctuation field admitting an effective coarse-grained description. In the present special-case analysis, the finite correlation time of the microscopic fluctuations is identified with the Planck time, \(t_{\rm Pl}\). In this way, the slow background-induced modulation of the variance discussed above is encoded in a time-dependent coarse-grained amplitude, while the rapid microscopic fluctuations are characterized by the Planck-scale correlation time. Concretely, we assume that the two-point coarse-grained correlation function takes the form
	\begin{equation}
		C(t,t')
		\equiv
		\langle b(t)b(t')\rangle
		=
		\sigma_b(t)\sigma_b(t')
		K\!\left(\frac{t-t'}{t_{\rm Pl}}\right),
		\label{eq:colored-noise-covariance}
	\end{equation}
	where \(K\) is a smooth even kernel satisfying \(K(0)=1\), and \(\sigma_b(t)\) is a slowly varying coarse-grained amplitude obeying
	\begin{equation}
		t_{\rm Pl}\left|\frac{\dot{\sigma}_b}{\sigma_b}\right|\ll1,
		\qquad
		t_{\rm Pl}^2\left|\frac{\ddot{\sigma}_b}{\sigma_b}\right|\ll1.
		\label{eq:slow-noise-amplitude}
	\end{equation}
	Thus, the correlation decays on the short Planckian timescale, whereas the coarse-grained variance \(\langle b^2(t)\rangle=\sigma_b^2(t)\) may evolve smoothly on the cosmological background timescale. For definiteness, we choose the Gaussian kernel
	\begin{equation}
		K(u)=\exp\!\left(-\frac{u^2}{2}\right).
		\label{eq:gaussian-noise-kernel}
	\end{equation}
	This choice provides a well-defined stochastic process with finite derivative moments, in contrast to ideal white noise. Moreover, because the temporal variation of the kernel is controlled by the short Planck-scale correlation time, the derivative variance scales as
	\begin{equation}
		\langle \dot b^2(t)\rangle
		\sim
		\frac{\sigma_b^2(t)}{t_{\rm Pl}^2}
		=
		\frac{\langle b^2(t)\rangle}{t_{\rm Pl}^2},
		\label{eq:db2-scaling-special-cases}
	\end{equation}
	up to a numerical factor of order unity fixed by the detailed shape of \(K\).

A useful model-independent measure of perturbative control is the
rms conformal amplitude
\begin{equation}
	\delta_{\rm rms}(R_h)
	\equiv
	\sqrt{\langle (\epsilon b)^2\rangle}
	=
	|\epsilon|\sqrt{\langle b^2\rangle},
	\label{eq:rms-noise-amplitude}
\end{equation}
which must satisfy
\begin{equation}
	\delta_{\rm rms}(R_h)\ll 1
	\label{eq:noise-perturbativity-condition}
\end{equation}
for the stochastic treatment to remain perturbative. Since
\begin{equation}
	S_{BH}(R_h)=\frac{\pi R_h^2}{G},
\end{equation}
a smaller horizon corresponds to a smaller horizon entropy.
Therefore, present-day smallness of $\mathcal{O}(\epsilon b)$ does
not by itself guarantee perturbativity at early times; the
small-$S_{BH}$ behavior must be assessed separately for each
entropy model.
\subsection{R\'enyi entropy}
For a discrete probability distribution
$P = (p_1,p_2,\dots,p_n)$ satisfying $\sum_{i=1}^n p_i = 1$,
the R\'enyi entropy of order $\zeta$ is defined as
\begin{equation}
    S_{\zeta}
    =
    \frac{1}{1-\zeta}
    \ln\!\left(\sum_{i=1}^n p_i^{\zeta}\right),
    \label{eq:Renyi-general}
\end{equation}
where $\zeta$ is a real parameter characterizing the degree of nonextensivity.
In the limit $\zeta \to 1$, the R\'enyi entropy reduces to the standard Shannon entropy,
\begin{equation}
    S_{\rm Sh}
    =
    -\sum_{i=1}^n p_i \ln p_i.
\end{equation}
In the context of black hole thermodynamics, the R\'enyi entropy
provides a logarithmic deformation of the Bekenstein--Hawking area
law \cite{komatsu2017,moradpour2017}
\begin{equation}
    S_R
    =
    \frac{1}{\lambda}
    \ln\!\left(1+\lambda S_{BH}\right),
    \label{eq:Renyi-horizon}
\end{equation}
where $\lambda$ is the R\'enyi deformation parameter. In the limit
$\lambda \to 0$, one recovers the standard area law $S_R \to
S_{BH}$. At first order in $\lambda$, using
Eq.~\eqref{eq:Friedmann-general-mod}, the modified Friedmann
equation reads as
\begin{equation}
    \frac{8\pi G}{3}\rho
    =
    \frac{1}{R_h^2}
    -\frac{\pi\lambda}{G}\ln\!\left(\frac{G}{\pi R_h^2}\right).
\end{equation}
Using Eq.~\eqref{eq:match-noise-approx}, one may match
\begin{equation}
    \epsilon^2=4\pi\lambda, \qquad
    \langle b^2\rangle
    = \ln\!\left(S_{BH}(R_h)\right),
    \label{eq:Noise-Renyi}
\end{equation}
where $S_{BH}(R_h)$ denotes the Bekenstein-Hawking entropy of the
horizon $R_h$. This matching is restricted to the real stochastic sector, so that $\lambda\ge 0$; negative $\lambda$ would fall outside the present real-noise correspondence. The corresponding noise therefore exhibits a standard
deviation
\begin{equation}
    \Delta b=\sqrt{\ln(S_{BH}(R_h))},
\end{equation}
which exhibits only a weak dependence on the horizon scale $R_h$.
This behavior implies that the width of the stochastic
fluctuations grows extremely slowly as the Universe expands. Using
phenomenological constraints on $\lambda$ obtained in
\cite{sheykhi2025b}, one finds
\begin{equation}
    \mathcal{O}(\delta_{\rm rms}^{\rm R})= 10^{-42} \sqrt{\ln(S_{BH}(R_h))}\sim 10^{-40},
\end{equation}
for the present Universe. Such an extremely small value is fully consistent with treating
the conformal noise as a perturbative correction to the FRW
background. Since the associated rms amplitude becomes smaller for
smaller horizon entropy, the stochastic contribution remains
perturbatively suppressed also at earlier times. It is therefore
expected to have had no significant impact on the cosmological
evolution, including in the early-Universe regime, within the
domain of validity of the present effective framework.

\subsection{(Dual) Kaniadakis entropy}
The Kaniadakis entropy is defined as
\begin{equation}
    S_{\kappa}=\frac{1}{\kappa}\sinh(\kappa S_{BH}),
\end{equation}
and produces, at leading order, a correction to the Friedmann equation of the form
\begin{equation}
    \frac{8\pi G}{3}\rho
    =\frac{1}{R_h^2}-\frac{\kappa^2\pi^2}{2 G^2} R_h^2.
\end{equation}
Since the noise-induced contribution in
Eq.~\eqref{eq:modifiedFriedmann-2} is positive definite, this
correction carries the opposite sign. Nevertheless, one may use
the dual Kaniadakis entropy with parameter
$\kappa^{*2}=-\kappa^2$. Hence the modified Friedmann equation
becomes
\begin{equation}
    \frac{8\pi G}{3}\rho
    =\frac{1}{R_h^2}+\frac{\kappa^{*2}\pi^2}{2 G^2} R_h^2.
\end{equation}
Using Eq.~\eqref{eq:match-noise-approx}, one may match
\begin{equation}
    \epsilon^2=2\pi\kappa^{*2}, \qquad
    \langle b^2\rangle
    = S_{BH}(R_h).
    \label{eq:Noise-Kaniadakis}
\end{equation}
The corresponding noise exhibits a standard deviation
\begin{equation}
    \Delta b=\sqrt{S_{BH}(R_h)}.
\end{equation}
The associated noise distribution may therefore be regarded as
effectively scale-free, since its dispersion is determined
directly by the horizon size. Consequently, the standard deviation
grows linearly with $R_h$ as the Universe expands. Using
phenomenological bounds on $\kappa^*$ from \cite{sheykhi2025a},
one obtains
\begin{equation}
    \mathcal{O}(\delta_{\rm rms}^{\rm DK})= 10^{-69} \sqrt{S_{BH}(R_h)}\sim 10^{-8},
\end{equation}
for the present Universe. This remains sufficiently small to
justify treating the conformal noise as a perturbative correction
to the FRW background. If the deformation parameter is assumed to be
scale independent, the conformal-noise amplitude decreases toward
earlier times as the horizon entropy becomes smaller, so its
impact on the past cosmological evolution is expected to be
less significant than at present. On the other hand, the
stochastic contribution grows as the universe expands. A simple
extrapolation of the present estimate therefore suggests that if
the horizon size were to become of order $10^8$ times larger than
its current value, then $\delta_{\rm rms}^{\rm DK}$ would become
of order unity and the perturbative treatment would cease to be
valid. This should be understood as an indication of the
eventual breakdown of the effective description in the far future.

\subsection{Barrow entropy}
Barrow entropy modifies the area law through a fractal deformation
of the horizon surface,
\begin{equation}
    S_B=\left(\frac{A}{4G}\right)^{1+\Delta/2},
\end{equation}
where $\Delta$ quantifies the deviation from a smooth horizon geometry.
To first order in $\Delta$, the corresponding Friedmann equation becomes
\begin{equation}
    \frac{8\pi G}{3}\rho
    =
    \frac{1}{R_h^2}
    +\Delta\left(\frac{1}{R_h^2}+\frac{\ln(\pi R_h^2/G)}{2R_h^2}\right).
\end{equation}
Therefore, using Eq.~\eqref{eq:match-noise-approx}, one may match this correction through
\begin{equation}
    \epsilon^2=4\pi\Delta, \qquad
    \langle b^2\rangle
    = \frac{1}{\pi R_h^2/G}+\frac{\ln(\pi R_h^2/G)}{2\pi R_h^2/G}\sim \frac{1}{S_{BH}(R_h)}.
    \label{eq:Noise-Barrow}
\end{equation}
This matching is likewise understood in the real stochastic sector, requiring $\Delta\ge 0$; negative $\Delta$ is not considered in the present correspondence. The standard deviation therefore scales as
\begin{equation}
    \Delta b\sim\frac{1}{\sqrt{S_{BH}(R_h)}},
\end{equation}
which decreases with increasing horizon radius. Thus, in an
expanding Universe, the conformal noise distribution becomes
progressively more sharply peaked around zero. Using observational
constraints $\Delta=\mathcal{O}(10^{-4})$ \cite{barrow2021}, one
finds
\begin{equation}
	\label{eq:deltaB-1}
    \mathcal{O}(\delta_{\rm rms}^{\rm B})= \frac{10^{-1}}{\sqrt{S_{BH}(R_h)}}\sim 10^{-62},
\end{equation}
for the present Universe. Such an extremely small amplitude is
fully consistent with treating the conformal noise as a
perturbative correction to the FRW background.

An important feature of the Barrow case is that, unlike
	scenarios in which the stochastic sector becomes progressively less
	relevant toward the past, the Barrow-induced contribution is expected
	to be less suppressed in the early universe and may therefore acquire
	genuine cosmological significance there. To estimate
	when this contribution may become cosmologically relevant, it is
	natural to compare its rms amplitude with the observed order of
	magnitude of primordial cosmological fluctuations, namely
	$\sim10^{-5}$, as inferred from CMB observations. In this sense, the condition
	$\delta_{\rm rms}^{\rm B}\sim10^{-5}$ is a phenomenologically motivated
	benchmark for the epoch at which the Barrow-induced stochastic sector
	may begin to have non-negligible cosmological impact. One finds from
	Eq.~\eqref{eq:deltaB-1} that this condition is reached when
	$S_{BH}\sim10^{8}$. At the order-of-magnitude level, this corresponds
	to a horizon radius $R_h\sim10^{4}l_p$. Interpreting the horizon scale
	as $R_h\sim H^{-1}$ then gives an expansion rate of order
	$H\sim10^{-4}M_p$, which in a standard high-energy radiation
	background points to a characteristic energy scale roughly in the
	range $10^{16}$--$10^{17}\,\mathrm{GeV}$, i.e. near the conventional
	GUT/primordial-inflationary regime.

This estimate is physically noteworthy because it shows that
	the Barrow-induced stochastic conformal sector can become
	cosmologically non-negligible already in a very early, yet still
	perturbative, regime. In the present formulation, however, the
	stochastic conformal mode depends only on cosmic time, so its effect
	is homogeneous and should be interpreted primarily as a stochastic
	modification of the background evolution rather than as a direct
	source of genuinely spatial primordial inhomogeneities. From this
	perspective, the Barrow profile is not merely a negligible late-time
	correction, but may instead encode an effective high-curvature
	stochastic contribution to the early-universe background dynamics.
	This makes the Barrow case particularly interesting as a
	phenomenological probe of how generalized entropy deformations may
	manifest themselves in the pre-inflationary or early-universe
	regime.

According to Eq.~\eqref{eq:deltaB-1}, a formal breakdown of
	the perturbative description, $\delta_{\rm rms}^{\rm B}\sim1$, would
	occur near $S_{BH}\sim10^{-2}$, namely deep in the sub-unity
	entropy regime where the semiclassical interpretation itself is
	already expected to fail. Therefore, throughout the physically
	relevant large-entropy domain, the Barrow correspondence remains
	perturbatively meaningful while still allowing a much stronger
	early-time stochastic imprint than at present.

\subsection{Logarithmic and inverse-area corrections}
Quantum gravity approaches often predict subleading corrections to
the Bekenstein--Hawking entropy beyond the leading area term. A
widely studied parametrization takes the form
\begin{equation}
    S_{QG}=\frac{A}{4G}
    +\beta\ln\!\left(\frac{A}{4G}\right)
    +\gamma\frac{4G}{A}
    +\cdots,
    \label{eq:log-inv-entropy}
\end{equation}
where $\beta$ and $\gamma$ are dimensionless coefficients expected to be of $\mathcal{O}(10^{-1}\!-\!10^{0})$ \cite{xiao2022,cai2008}. Using the general entropy-modified Friedmann relation \eqref{eq:Friedmann-general-mod}, the corresponding Friedmann equation may be written as
\begin{equation}
    \frac{8\pi G}{3}\rho
    =
    \frac{1}{R_h^2}
    +\frac{\beta}{2\pi}\frac{G}{R_h^4}-\frac{\gamma}{3\pi^2}\frac{G^2}{R_h^6}.
    \label{eq:friedmann-log-inv}
\end{equation}
Comparing Eq.~\eqref{eq:friedmann-log-inv} with the noise-corrected Friedmann equation in the fast-oscillation regime, Eq.~\eqref{eq:modifiedFriedmann-2}, suggests that these
quantum-induced entropy corrections can likewise be reproduced within the conformal noise framework. In particular, one may match
\begin{equation}
    \epsilon^2 = \frac{2}{\pi}\beta,
    \qquad
    \langle b^2\rangle = \frac{G^2}{R_h^4}\sim \frac{1}{S^2_{BH}(R_h)},
    \label{eq:Noise-Log}
\end{equation}
for the logarithmic correction term, while the higher-order inverse-area contribution corresponds to a stronger suppression
\begin{equation}
    \epsilon^2 = -\frac{4}{3\pi^2}\gamma,
    \qquad
    \langle b^2\rangle = \frac{G^3}{R_h^6}\sim \frac{1}{S^3_{BH}(R_h)}.
    \label{eq:Noise-InvArea}
\end{equation}
This identification is valid for $\beta \geq 0$ (logarithmic) and $\gamma \leq 0$ (inverse-area), ensuring a non-negative variance; sign-incompatible regimes lie outside the scope of the present framework. The corresponding standard deviations and present-day orders of magnitude for the noise profiles become
\begin{equation}
    \Delta b\sim\frac{1}{S_{BH}(R_h)},\qquad\mathcal{O}(\delta_{\rm rms}^{\rm log})= \frac{1}{S_{BH}(R_h)}\sim 10^{-122},
\end{equation}
for the logarithmic correction, and
\begin{equation}
    \Delta b\sim\frac{1}{S^{3/2}_{BH}(R_h)},\qquad\mathcal{O}(\delta_{\rm rms}^{\rm Inv})= \frac{1}{S^{3/2}_{BH}(R_h)}\sim 10^{-183},
\end{equation}
for the inverse-area contribution. These extremely small values
indicate that such quantum-induced corrections correspond, within
the conformal noise picture, to highly suppressed stochastic
fluctuations of the spacetime geometry. A noteworthy feature of both the logarithmic and inverse-area
	corrections is that, although their associated stochastic amplitudes
	increase toward earlier times as the horizon entropy decreases, they
	remain much more strongly suppressed than in the Barrow case. In
	particular, within
	the entire large-entropy semiclassical regime, these contributions are
	expected to remain extremely small and therefore cosmologically
	subleading.

\subsection{MOND-inspired hypergeometric entropy}
A more general class of entropy deformations arises from the
MOND-inspired construction of \cite{sheykhi2025}, where the
entropy associated with the horizon takes the hypergeometric form
\begin{equation}
    S_\text{MOND}=\frac{A}{4G}\;
    {}_2F_1\!\left(\frac1\alpha,\frac1\alpha,\frac{\alpha+1}{\alpha},-
    \left(\frac{\gamma A}{4G}\right)^\alpha\right),
\end{equation}
with $\alpha>0$ being a free parameter. This construction interpolates between
several known entropy modifications and reproduces the R\'enyi and Kaniadakis
cases for the particular choices $\alpha=1$ and $\alpha=2$, respectively. For small deformation parameter $\gamma$, the entropy admits the expansion
\begin{equation}
    S_\text{MOND}=\frac{A}{4G}
    \left[
    1-\eta\left(\frac{A}{4G}\right)^\alpha
    +\cdots
    \right],
    \label{eq:MOND-expansion}
\end{equation}
where
\begin{equation}
    \eta=\frac{\gamma^\alpha}{\alpha(\alpha+1)}\ll1.
\end{equation}
Substituting this entropy correction into the general modified
Friedmann relation \eqref{eq:Friedmann-general-mod} yields
\begin{equation}
    \frac{8\pi G}{3}\rho
    =
    \frac{1}{R_h^2}
    +\frac{\alpha+1}{\alpha-1}\frac{\pi}{G}\eta \left(\frac{\pi R_h^2}{G}\right)^{\alpha-1}.
\end{equation}
Comparing with the noise-corrected Friedmann equation in the fast-oscillation
regime, Eq.~\eqref{eq:modifiedFriedmann-2}, one identifies
\begin{equation}
    \epsilon^2 = \frac{\alpha+1}{\alpha-1}4\pi\eta,
    \qquad
    \langle b^2\rangle
    = S_{BH}^{\alpha-1}(R_h).
    \label{eq:Noise-MOND}
\end{equation}
This identification requires $\alpha > 1$ to ensure a non-negative variance; the sector $\alpha < 1$ requires an opposite sign for the variance and is thus excluded from the present analysis.

An important feature of the MOND-inspired realization for
	$\alpha>1$ is that the corresponding
	stochastic amplitude becomes progressively smaller toward earlier
	times. Therefore, throughout the semiclassical large-entropy regime, the MOND-inspired stochastic
	contribution is expected to remain perturbatively small in the early
	universe and to have no significant impact on the
	cosmological evolution there.

\begin{table*}[t]
	\centering
	\begin{tabular}{c c c c}
		\hline
		Entropy model
		& Variance scaling
		& Present $\delta_{\rm rms}$
		& Early-time behavior \\
		\hline
		
		R\'enyi
		& $\ln S_{BH}$
		& $\sim 10^{-40}$
		& Further suppressed for smaller $S_{BH}$ \\
		
		Dual Kaniadakis
		& $S_{BH}$
		& $\sim 10^{-8}$
		& Decreases toward the past; possible future breakdown \\
		
		Barrow
		& $S_{BH}^{-1}$
		& $\sim 10^{-62}$
		& Enhanced toward the past; potentially relevant at high curvature \\
		
		Logarithmic
		& $S_{BH}^{-2}$
		& $\sim 10^{-122}$
		& Enhanced toward the past but still strongly suppressed \\
		
		Inverse-area
		& $S_{BH}^{-3}$
		& $\sim 10^{-183}$
		& Enhanced toward the past but ultra-suppressed \\
		
		MOND-inspired, $\alpha>1$
		& $S_{BH}^{\alpha-1}$
		& model dependent
		& Decreases toward the past \\
		
		\hline
	\end{tabular}
	\caption{Representative generalized entropy corrections reproduced within the conformal noise framework. The physically relevant perturbative quantity is the rms conformal amplitude $\delta_{\rm rms}=|\epsilon|\sqrt{\langle b^2\rangle}$.}
	\label{tab:entropy-noise}
\end{table*}
\subsection{Physical Interpretation and Comparison}
The examples discussed above show that several generalized entropy
corrections can be reproduced at the level of the effective Friedmann
equation by assigning suitable horizon-dependent statistical
properties to the conformal noise. This correspondence should,
however, be interpreted with some care. The variance scalings listed
in Table~\ref{tab:entropy-noise} do not by themselves determine the
physical size of the metric fluctuation. The perturbative quantity is
instead the rms conformal amplitude
$\delta_{\rm rms}=|\epsilon|\sqrt{\langle b^2\rangle}$, which depends
both on the variance profile and on the entropy-deformation parameter. In this respect, the R\'enyi case is among the mildest realizations:
its variance grows only logarithmically,
$\langle b^2\rangle\propto \ln S_{BH}$, so that the corresponding rms amplitude decreases toward
earlier times. Therefore, throughout the physically relevant
large-entropy semiclassical regime, the R\'enyi contribution remains
safely perturbative and is best viewed as a weak stochastic dressing
of the background rather than a source of significant
early-universe effects.

This distinction is particularly important for the dual Kaniadakis
case. There, the variance scales as $\langle b^2\rangle\propto
S_{BH}$ and is therefore enormous for the present cosmological
horizon. The stochastic correction nevertheless remains small because
the phenomenological bound on the deformation parameter makes
$|\epsilon|$ extremely small. Thus, the present-day perturbativity of
this sector is not a consequence of a small variance, but of a highly
suppressed coupling between the conformal noise and the background
geometry. If this coupling is assumed to be scale independent, the
dual Kaniadakis contribution becomes less important toward the past,
but may eventually become non-perturbative in a far-future regime with
a much larger horizon entropy.

The Barrow, logarithmic, and inverse-area cases display the opposite
type of behavior: their variances decrease with increasing
$S_{BH}$ and are therefore extremely suppressed for the present
horizon. Toward earlier times these amplitudes grow as the horizon
entropy decreases. Among them, the Barrow profile is the most
potentially relevant, since its rms amplitude scales only as
$S_{BH}^{-1/2}$ and can become non-negligible in a high-curvature yet
still semiclassical regime. By contrast, the logarithmic and
inverse-area corrections scale as higher inverse powers of
$S_{BH}$ and remain much more strongly suppressed throughout the
large-entropy domain. They are therefore best interpreted as
subleading stochastic dressings of the homogeneous background rather
than as sources of sizeable cosmological effects.

The MOND-inspired hypergeometric model provides a continuous family
of power-law profiles. In the real-noise sector considered here,
$\alpha>1$ leads to $\langle b^2\rangle\propto S_{BH}^{\alpha-1}$, so
the associated stochastic amplitude decreases toward earlier times.
Consequently, this class behaves qualitatively like the dual
Kaniadakis case rather than the Barrow case: it is not expected to
generate enhanced early-universe stochastic effects within the
semiclassical regime, although a future breakdown of perturbativity
may occur if the horizon entropy becomes sufficiently large.

Finally, the noise profiles obtained here should not be regarded as
microscopically derived stochastic kernels. At present, no unique
underlying quantum-gravity theory is known to predict these precise
horizon-dependent variances for a time-dependent conformal mode. The
models should instead be viewed as phenomenological realizations of
entropy-corrected Friedmann dynamics within a stochastic geometric
language. Their physical plausibility is therefore assessed
operationally: the correspondence is meaningful only in regimes where
the rms conformal amplitude remains perturbative and where the
semiclassical notion of horizon entropy is still applicable.
\section{Conclusions}  \label{Sec:Concs}
In this work, we have explored an effective framework where modifications to the Friedmann equations arise from the coarse-graining of stochastic conformal fluctuations of the spacetime metric. By considering a family of stochastic configurations around a spatially flat FRW background with vanishing cosmological constant, we showed that the nonlinearities in Einstein's equations, when averaged to second order in the fluctuation amplitude, generate a correction term that can be effectively matched to various generalized entropy formalisms.

Our results suggest a phenomenological correspondence between the statistical properties of metric noise and the functional forms of the modified Friedmann equations within the restricted background studied here. Rather than providing a first-principles derivation of entropy laws, this approach provides a consistent way to describe the macroscopic effects of unresolved microscopic degrees of freedom through effective modifications of horizon thermodynamics. Specifically, we have shown that, by adopting a variance profile for the noise as a matching ansatz, one can obtain an effective correspondence with the dynamics associated with R\'enyi, (dual) Kaniadakis, Barrow, logarithmic, and inverse-area entropy models in the spatially flat case with vanishing cosmological constant. This illustrates the versatility of the stochastic framework in capturing diverse modifications to the standard area law within a unified effective picture.

Future work could extend this analysis beyond conformal perturbations to include more general stochastic fluctuations, explore the implications for early-Universe observables such as primordial nucleosynthesis and the cosmic microwave background, and investigate whether specific noise correlations might lead to distinctive observational signatures capable of discriminating between different entropy models. Additionally, establishing a more direct connection between the statistical properties of the noise and underlying quantum-gravity frameworks remains an important direction for future research.
\acknowledgments{We thank Shiraz University Research Council. We also acknowledge the use of artificial intelligence tools for assistance with improving the presentation of the manuscript.}


\begin{thebibliography}{10}

            \bibitem{jacobson1995}
            T.~Jacobson, \emph{Thermodynamics of {{Spacetime}}: {{The Einstein Equation}}
                of {{State}}}, \href{https://doi.org/10.1103/PhysRevLett.75.1260}{\emph{Phys.
                    Rev. Lett.} {\bfseries 75} (1995) 1260}
                [\href{https://arxiv.org/abs/gr-qc/9504004}{{\ttfamily arXiv:gr-qc/9504004}}].

            \bibitem{eling2006}
            C.~Eling, R.~Guedens and T.~Jacobson, \emph{Nonequilibrium {{Thermodynamics}}
                of {{Spacetime}}},
            \href{https://doi.org/10.1103/PhysRevLett.96.121301}{\emph{Phys. Rev. Lett.}
                {\bfseries 96} (2006) 121301}
            [\href{https://arxiv.org/abs/gr-qc/0602001}{{\ttfamily arXiv:gr-qc/0602001}}].

            \bibitem{padmanabhan2005}
            T.~Padmanabhan, \emph{Gravity and the thermodynamics of horizons},
            \href{https://doi.org/10.1016/j.physrep.2004.10.003}{\emph{Phys. Reports}
                {\bfseries 406} (2005) 49}
            [\href{https://arxiv.org/abs/gr-qc/0311036}{{\ttfamily
                    arXiv:gr-qc/0311036}}].

            \bibitem{paranjape2006}
            A.~Paranjape, S.~Sarkar and T.~Padmanabhan, \emph{Thermodynamic route to field
                equations in {{Lanczos-Lovelock}} gravity},
            \href{https://doi.org/10.1103/PhysRevD.74.104015}{\emph{Phys. Rev. D}
                {\bfseries 74} (2006) 104015}
            [\href{https://arxiv.org/abs/hep-th/0607240}{{\ttfamily
                    arXiv:hep-th/0607240}}].

            \bibitem{akbar2006}
            M.~Akbar and R.-G.~Cai, \emph{Friedmann equations of {{FRW}} universe in
                scalar--tensor gravity, f(R) gravity and first law of thermodynamics},
            \href{https://doi.org/10.1016/j.physletb.2006.02.035}{\emph{Phys.
            Lett. B} {\bfseries 635} (2006) 7}
        [\href{https://arxiv.org/abs/hep-th/0602156}{{\ttfamily
                arXiv:hep-th/0602156}}].

            \bibitem{padmanabhan2010a}
            T.~Padmanabhan, \emph{Thermodynamical aspects of gravity: New insights},
            \href{https://doi.org/10.1088/0034-4885/73/4/046901}{\emph{Rep. Prog. Phys.}
                {\bfseries 73} (2010) 046901}
            [\href{https://arxiv.org/abs/0911.5004}{{\ttfamily
                    arXiv:0911.5004}}].

            \bibitem{wang2006}
            B.~Wang, Y.~Gong and E.~Abdalla, \emph{Thermodynamics of an accelerated
                expanding universe},
            \href{https://doi.org/10.1103/PhysRevD.74.083520}{\emph{Phys. Rev. D}
                {\bfseries 74} (2006) 083520}
            [\href{https://arxiv.org/abs/gr-qc/0511051}{{\ttfamily
                    arXiv:gr-qc/0511051}}].

            \bibitem{cai2005}
            R.-G.~Cai and S.P.~Kim, \emph{First {{Law}} of {{Thermodynamics}} and
                {{Friedmann Equations}} of {{Friedmann}}--{{Robertson}}--{{Walker
                        Universe}}}, \href{https://doi.org/10.1088/1126-6708/2005/02/050}{\emph{J.
                    High Energy Phys.} {\bfseries02} (2005) 050}
                [\href{https://arxiv.org/abs/hep-th/0501055}{{\ttfamily
                        arXiv:hep-th/0501055}}].

            \bibitem{akbar2007}
            M.~Akbar and R.-G.~Cai, \emph{Thermodynamic behavior of the {{Friedmann}}
                equation at the apparent horizon of the {{FRW}} universe},
            \href{https://doi.org/10.1103/PhysRevD.75.084003}{\emph{Phys. Rev. D}
                {\bfseries 75} (2007) 084003}
            [\href{https://arxiv.org/abs/hep-th/0609128}{{\ttfamily
                    arXiv:hep-th/0609128}}].

            \bibitem{sheykhi2007a}
            A.~Sheykhi, B.~Wang and R.-G.~Cai, \emph{Deep connection between thermodynamics
                and gravity in {{Gauss-Bonnet}} braneworlds},
            \href{https://doi.org/10.1103/PhysRevD.76.023515}{\emph{Phys. Rev. D}
                {\bfseries 76} (2007) 023515}
            [\href{https://arxiv.org/abs/hep-th/0701261}{{\ttfamily
                    arXiv:hep-th/0701261}}].

            \bibitem{sheykhi2010a}
            A.~Sheykhi, \emph{Thermodynamics of interacting holographic dark energy with
                the apparent horizon as an {{IR}} cutoff},
            \href{https://doi.org/10.1088/0264-9381/27/2/025007}{\emph{Class. Quantum
                    Grav.} {\bfseries 27} (2010) 025007}
                [\href{https://arxiv.org/abs/0910.0510}{{\ttfamily
                        arXiv:0910.0510}}].

            \bibitem{sheykhi2010b}
            A.~Sheykhi, \emph{Thermodynamics of apparent horizon and modified {{Friedmann}}
                equations}, \href{https://doi.org/10.1140/epjc/s10052-010-1372-9}{\emph{Eur.
                    Phys. J. C} {\bfseries 69} (2010) 265}
                [\href{https://arxiv.org/abs/1012.0383}{{\ttfamily
                        arXiv:1012.0383}}].

            \bibitem{verlinde2011}
            E.~Verlinde, \emph{On the origin of gravity and the laws of {{Newton}}},
            \href{https://doi.org/10.1007/JHEP04(2011)029}{\emph{J. High Energ. Phys.}
                {\bfseries 04} (2011) 29}
            [\href{https://arxiv.org/abs/1001.0785}{{\ttfamily
                    arXiv:1001.0785}}].

            \bibitem{cai2010}
            R.-G.~Cai, L.-M.~Cao and N.~Ohta, \emph{Friedmann equations from entropic
                force}, \href{https://doi.org/10.1103/PhysRevD.81.061501}{\emph{Phys. Rev. D}
                {\bfseries 81} (2010) 061501}
            [\href{https://arxiv.org/abs/1001.3470}{{\ttfamily
                    arXiv:1001.3470}}].

            \bibitem{sheykhi2010}
            A.~Sheykhi, \emph{Entropic corrections to {{Friedmann}} equations},
            \href{https://doi.org/10.1103/PhysRevD.81.104011}{\emph{Phys. Rev. D}
                {\bfseries 81} (2010) 104011}
            [\href{https://arxiv.org/abs/1004.0627}{{\ttfamily
                arXiv:1004.0627}}].

            \bibitem{visser2011}
            M.~Visser, \emph{Conservative entropic forces},
            \href{https://doi.org/10.1007/JHEP10(2011)140}{\emph{J. High Energ. Phys.}
                {\bfseries 1110} (2011) 140}
            [\href{https://arxiv.org/abs/1108.5240}{{\ttfamily
                    arXiv:1108.5240}}].

            \bibitem{sheykhi2013}
            A.~Sheykhi, H.~Moradpour and N.~Riazi, \emph{Lovelock gravity from entropic
                force},
                \href{https://doi.org/10.1007/s10714-013-1509-x}{\emph{Gen.
                Relativ. Gravit.} {\bfseries 45} (2013) 1033}
            [\href{https://arxiv.org/abs/1109.3631}{{\ttfamily
                    arXiv:1109.3631}}].

            \bibitem{padmanabhan2012}
            T.~Padmanabhan, \emph{Emergence and {{Expansion}} of {{Cosmic Space}} as due to
                the {{Quest}} for {{Holographic Equipartition}}},
            [\href{https://arxiv.org/abs/1206.4916}{{\ttfamily
                    arXiv:1206.4916}}].

            \bibitem{cai2012}
            R.-G.~Cai, \emph{Emergence of space and spacetime dynamics of
                {{Friedmann-Robertson-Walker}} universe},
            \href{https://doi.org/10.1007/JHEP11(2012)016}{\emph{J. High Energ. Phys.}
                {\bfseries 11} (2012) 016}
            [\href{https://arxiv.org/abs/1207.0622}{{\ttfamily
                    arXiv:1207.0622}}].

            \bibitem{yang2012}
            K.~Yang, Y.-X.~Liu and Y.-Q.~Wang, \emph{Emergence of cosmic space and the
                generalized holographic equipartition},
            \href{https://doi.org/10.1103/PhysRevD.86.104013}{\emph{Phys. Rev. D}
                {\bfseries 86} (2012) 104013}
            [\href{https://arxiv.org/abs/1207.3515}{{\ttfamily
                    arXiv:1207.3515}}].

            \bibitem{sheykhi2013a}
            A.~Sheykhi, \emph{Friedmann equations from emergence of cosmic space},
            \href{https://doi.org/10.1103/PhysRevD.87.061501}{\emph{Phys. Rev. D}
                {\bfseries 87} (2013) 061501}
            [\href{https://arxiv.org/abs/1304.3054}{{\ttfamily
                    arXiv:1304.3054}}].

            \bibitem{nojiri2019}
            S.~Nojiri, S.D.~Odintsov and E.N.~Saridakis, \emph{Modified cosmology from
                extended entropy with varying exponent},
            \href{https://doi.org/10.1140/epjc/s10052-019-6740-5}{\emph{Eur. Phys. J. C}
                {\bfseries 79} (2019) 242}
            [\href{https://arxiv.org/abs/1903.03098}{{\ttfamily
                    arXiv:1903.03098}}].

            \bibitem{nojiri2022}
            S.~Nojiri, S.D.~Odintsov and T.~Paul, \emph{Early and late universe holographic
                cosmology from a new generalized entropy},
            \href{https://doi.org/10.1016/j.physletb.2022.137189}{\emph{Phys.
            Lett. B} {\bfseries 831} (2022) 137189}
        [\href{https://arxiv.org/abs/2205.08876}{{\ttfamily
                arXiv:2205.08876}}].

            \bibitem{odintsov2023}
            S.D.~Odintsov and T.~Paul, \emph{A non-singular generalized entropy and its
                implications on bounce cosmology},
            \href{https://doi.org/10.1016/j.dark.2022.101159}{\emph{Phys. Dark
                    Univ.} {\bfseries 39} (2023) 101159}
                [\href{https://arxiv.org/abs/2212.05531}{{\ttfamily
                        arXiv:2212.05531}}].

            \bibitem{cai2008}
            R.-G.~Cai, L.-M.~Cao and Y.-P.~Hu, \emph{Corrected entropy-area relation and
                modified Friedmann equations},
            \href{https://doi.org/10.1088/1126-6708/2008/08/090}{\emph{J. High Energy
                    Phys.} {\bfseries 0808} (2008) 090}
                [\href{https://arxiv.org/abs/0807.1232}{{\ttfamily
                        arXiv:0807.1232}}].

            \bibitem{sheykhi2018}
            A.~Sheykhi, \emph{Modified {{Friedmann}} equations from {{Tsallis}} entropy},
            \href{https://doi.org/10.1016/j.physletb.2018.08.036}{\emph{Phys.
            Lett. B} {\bfseries 785} (2018) 118}
        [\href{https://arxiv.org/abs/1806.03996}{{\ttfamily
                arXiv:1806.03996}}].

            \bibitem{saridakis2020}
            E.N.~Saridakis, \emph{Modified cosmology through spacetime thermodynamics and
                {{Barrow}} horizon entropy},
            \href{https://doi.org/10.1088/1475-7516/2020/07/031}{\emph{J. Cosmol.
                    Astropart. Phys.} {\bfseries 07} (2020) 031}
                [\href{https://arxiv.org/abs/2006.01105}{{\ttfamily
                        arXiv:2006.01105}}].

            \bibitem{sheykhi2021}
            A.~Sheykhi, \emph{Barrow entropy corrections to {{Friedmann}} equations},
            \href{https://doi.org/10.1103/PhysRevD.103.123503}{\emph{Phys. Rev. D}
                {\bfseries 103} (2021) 123503}
                [\href{https://arxiv.org/abs/2102.06550}{{\ttfamily
                    arXiv:2102.06550}}].

            \bibitem{sheykhi2023}
            A.~Sheykhi, \emph{Modified cosmology through {{Barrow}} entropy},
            \href{https://doi.org/10.1103/PhysRevD.107.023505}{\emph{Phys. Rev. D}
                {\bfseries 107} (2023) 023505}
            [\href{https://arxiv.org/abs/2210.12525}{{\ttfamily
                    arXiv:2210.12525}}].

            \bibitem{tsallis1988}
            C.~Tsallis, \emph{Possible generalization of {{Boltzmann-Gibbs}} statistics},
            \href{https://doi.org/10.1007/BF01016429}{\emph{J. Stat. Phys.} {\bfseries 52}
                (1988) 479}.

            \bibitem{barrow2020}
            J.D.~Barrow, \emph{The area of a rough black hole},
            \href{https://doi.org/10.1016/j.physletb.2020.135643}{\emph{Phys.
            Lett. B} {\bfseries 808} (2020) 135643}
        [\href{https://arxiv.org/abs/2004.09444}{{\ttfamily
                arXiv:2004.09444}}].

            \bibitem{kaniadakis2002}
            G.~Kaniadakis, \emph{Statistical mechanics in the context of special
                relativity}, \href{https://doi.org/10.1103/PhysRevE.66.056125}{\emph{Phys.
                    Rev. E} {\bfseries 66} (2002) 056125}
                [\href{https://arxiv.org/abs/cond-mat/0210467}{{\ttfamily
                        arXiv:cond-mat/0210467}}].

            \bibitem{kaniadakis2005}
            G.~Kaniadakis, \emph{Statistical mechanics in the context of special
                relativity. {{II}}.},
            \href{https://doi.org/10.1103/PhysRevE.72.036108}{\emph{Phys. Rev. E}
                {\bfseries 72} (2005) 036108}
            [\href{https://arxiv.org/abs/cond-mat/0507311}{{\ttfamily
                    arXiv:cond-mat/0507311}}].

            \bibitem{lymperis2021}
            A.~Lymperis, S.~Basilakos and E.N.~Saridakis, \emph{Modified cosmology through
                {{Kaniadakis}} horizon entropy},
            \href{https://doi.org/10.1140/epjc/s10052-021-09852-9}{\emph{Eur. Phys. J. C}
                {\bfseries 81} (2021) 1037}
            [\href{https://arxiv.org/abs/2108.12366}{{\ttfamily
                    arXiv:2108.12366}}].

            \bibitem{hernandez-almada2022}
            A.~{Hern{\'a}ndez-Almada}, G.~Leon, J.~Maga{\~n}a, M.A.~{Garc{\'i}a-Aspeitia},
            V.~Motta, E.N.~Saridakis et~al., \emph{Observational constraints and
                dynamical analysis of {{Kaniadakis}} horizon-entropy cosmology},
            \href{https://doi.org/10.1093/mnras/stac795}{\emph{Mont. Not.
                    Royal Astro. Society} {\bfseries 512} (2022) 5122}
                [\href{https://arxiv.org/abs/2112.04615}{{\ttfamily
                        arXiv:2112.04615}}].

            \bibitem{drepanou2022}
            N.~Drepanou, A.~Lymperis, E.N.~Saridakis and K.~Yesmakhanova, \emph{Kaniadakis
                holographic dark energy and cosmology},
            \href{https://doi.org/10.1140/epjc/s10052-022-10415-9}{\emph{Eur. Phys. J. C}
                {\bfseries 82} (2022) 449}
            [\href{https://arxiv.org/abs/2109.09181}{{\ttfamily
                    arXiv:2109.09181}}].

            \bibitem{zangeneh2024}
            M.K.~Zangeneh and A.~Sheykhi, \emph{Modified cosmology through {{Kaniadakis}}
                entropy}, \href{https://doi.org/10.1142/S0217732324501384}{\emph{Mod. Phys.
                    Lett. A} {\bfseries 39} (2024) 2450138}
                [\href{https://arxiv.org/abs/2311.01969}{{\ttfamily
                        arXiv:2311.01969}}].

            \bibitem{sheykhi2024}
            A.~Sheykhi, \emph{Corrections to {{Friedmann}} equations inspired by
                {{Kaniadakis}} entropy},
            \href{https://doi.org/10.1016/j.physletb.2024.138495}{\emph{Phys.
            Lett. B} {\bfseries 850} (2024) 138495}
        [\href{https://arxiv.org/abs/2302.13012}{{\ttfamily
                arXiv:2302.13012}}].
            
            \bibitem{Dehyadegari:2026PLB}
            A.~Dehyadegari and A.~Sheykhi, \emph{Modified entropy from action principle},
            \href{https://doi.org/10.1016/j.physletb.2026.140751}{\emph{Phys. Lett. B} 
            	{\bfseries 880} (2026) 140751}
            [\href{https://arxiv.org/abs/2605.29814}{{\ttfamily arXiv:2605.29814}}].

            \bibitem{martin1999}
            R.~Mart{\'{\i}}n and E.~Verdaguer, \emph{On the semiclassical
                {{Einstein-Langevin}} equation},
            \href{https://doi.org/10.1016/S0370-2693(99)01068-0}{\emph{Phys. Lett. B}
                {\bfseries 465} (1999) 113}
            [\href{https://arxiv.org/abs/gr-qc/9811070}{{\ttfamily
                    arXiv:gr-qc/9811070}}].

            \bibitem{hayward1998}
            S.A.~Hayward, \emph{Unified first law of black-hole dynamics and relativistic
                thermodynamics},
            \href{https://doi.org/10.1088/0264-9381/15/10/017}{\emph{Class. Quantum
                    Grav.} {\bfseries 15} (1998) 3147}
                [\href{https://arxiv.org/abs/gr-qc/9710089}{{\ttfamily
                        arXiv:gr-qc/9710089}}].

            \bibitem{hayward1999}
            S.A.~Hayward, S.~Mukohyama and M.~Ashworth, \emph{Dynamic black-hole entropy},
            \href{https://doi.org/10.1016/S0375-9601(99)00225-X}{\emph{Phys. Lett. A}
                {\bfseries 256} (1999) 347}
            [\href{https://arxiv.org/abs/gr-qc/9810006}{{\ttfamily
                    arXiv:gr-qc/9810006}}].

            \bibitem{bak2000}
            D.~Bak and S.-J.~Rey, \emph{Cosmic holography{\textsuperscript{+}}},
            \href{https://doi.org/10.1088/0264-9381/17/15/101}{\emph{Class. Quantum
                    Grav.} {\bfseries 17} (2000) L83}
                [\href{https://arxiv.org/abs/hep-th/9902173}{{\ttfamily
                        arXiv:hep-th/9902173}}].

            \bibitem{sheykhi2025}
            A.~Sheykhi and L.~Liravi, \emph{{{MOND}} theory and thermodynamics of
                spacetime},
                \href{https://doi.org/10.1016/j.dark.2025.101967}{\emph{Phys.
                 Dark Univ.} {\bfseries 49} (2025) 101967}
             [\href{https://arxiv.org/abs/2510.14345}{{\ttfamily
                    arXiv:2510.14345}}].

            \bibitem{sheykhi2025a}
            A.~Sheykhi, A.S.~Sooraki and L.~Liravi, \emph{Big-{{Bang}} nucleosynthesis
                constraints on (dual) {{Kaniadakis}} cosmology},
            \href{https://doi.org/10.1103/fg96-fjnw}{\emph{Phys. Rev. D} {\bfseries 112}
                (2025) 103546}
            [\href{https://arxiv.org/abs/2504.21146}{{\ttfamily
                    arXiv:2504.21146}}].

            \bibitem{sheykhi2025b}
            A.~Sheykhi and A.S.~Sooraki, \emph{Constraints on {{R{\'e}nyi Entropy}} through
                {{Primordial Big-Bang Nucleosynthesis}} and {{Baryogenesis}}},
            [\href{https://arxiv.org/abs/2507.14250}{{\ttfamily
                    arXiv:2507.14250}}].

            \bibitem{komatsu2017}
            N.~Komatsu, \emph{Cosmological model from the holographic equipartition law
                with a modified {{R{\'e}nyi}} entropy},
            \href{https://doi.org/10.1140/epjc/s10052-017-4800-2}{\emph{Eur. Phys. J. C}
                {\bfseries 77} (2017) 229}
            [\href{https://arxiv.org/abs/1611.04084}{{\ttfamily
                    arXiv:1611.04084}}].

            \bibitem{moradpour2017}
            H.~Moradpour, A.~Bonilla, E.M.C.~Abreu and J.A.~Neto, \emph{Accelerated cosmos
                in a nonextensive setup},
            \href{https://doi.org/10.1103/PhysRevD.96.123504}{\emph{Phys. Rev. D}
                {\bfseries 96} (2017) 123504}
            [\href{https://arxiv.org/abs/1711.08338}{{\ttfamily
                    arXiv:1711.08338}}].

            \bibitem{barrow2021}
            J.D.~Barrow, S.~Basilakos and E.N.~Saridakis, \emph{Big {{Bang
                        Nucleosynthesis}} constraints on {{Barrow}} entropy},
            \href{https://doi.org/10.1016/j.physletb.2021.136134}{\emph{Phys.
            Lett. B} {\bfseries 815} (2021) 136134}
        [\href{https://arxiv.org/abs/2010.00986}{{\ttfamily
                arXiv:2010.00986}}].

            \bibitem{xiao2022}
            Y.~Xiao and Y.~Tian, \emph{Logarithmic correction to black hole entropy from
                the nonlocality of quantum gravity},
            \href{https://doi.org/10.1103/PhysRevD.105.044013}{\emph{Phys. Rev. D}
                {\bfseries 105} (2022) 044013}
            [\href{https://arxiv.org/abs/2104.14902}{{\ttfamily
                    arXiv:2104.14902}}].
\end{thebibliography}
\end{document}